“Pernae salito. In salem ponito totas.”[1]
Cato the Elder, *De Agricultura* (ch. 162)

# The Physics Behind Ham Production

**Sergey Parnovsky**
Department of Astrophysics, National Taras Shevchenko University of Kyiv, Observatorna St. 3, Kyiv, 04053, Ukraine

**Andrey Varlamov**
CNR-SPIN, Via del Fosso del Cavaliere 100, 00133 Rome, Italy
Istituto Lombardo “Accademia di Scienze e Lettere”, Via Borgonuovo 25, Milan, Italy

***Abstract***

*We analyze the physical mechanisms underlying the production of dry-cured hams, including prosciutto and jamón. Key processes such as osmosis and diffusion govern water removal and salt penetration, while enzymatic activity during long-term air curing develops flavor and texture without heat. The study quantifies diffusion depths, evaporation rates, and the effects of airflow and humidity on preventing defects like case hardening. By highlighting the physics behind salting and curing, the work provides a deeper understanding of how traditional techniques achieve safe, tender, and flavorful hams.*

## *From Prosciutto to Jamón: Europe’s Dry-Cured Ham Traditions*

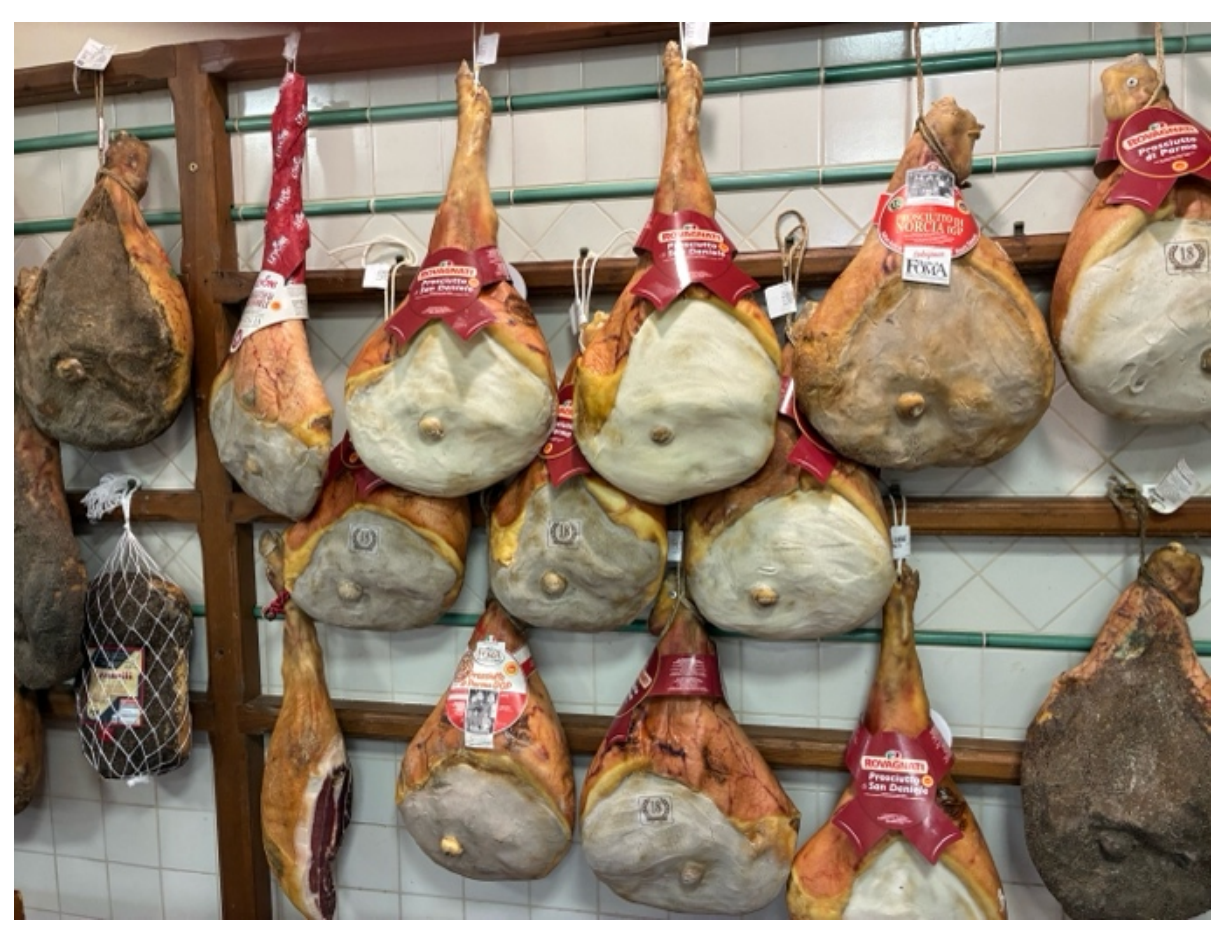

*Figure 1 A counter displaying a variety of prosciutto at the Lunghini store near Cortona in Tuscany.*

It is impossible to imagine Italian cuisine without prosciutto crudo, or Spanish cuisine without jamón. Both of these delicacies are similar in their method of preparation: after a relatively short period of dry salting (from two to four weeks), the meat is air-cured for several months or even years without undergoing any heat treatment. The result is a dry-cured ham with firm, dry meat that has a distinctive taste and aroma.

The varieties of the product differ in the way the animals are fed, their breeds, and the spices used in preparation. Jamón serrano is cured from the meat of white-breed swine, while jamón curado is made from the meat of black-breed swine. The most famous Italian variety of prosciutto, known as Prosciutto di Parma (Parma ham), is made from pork obtained from animals fed on Parmesan whey and seasoned only with sea salt. The animals used for the production of Prosciutto di Modena must belong to white breeds and may originate only from ten regions of Central and Northern Italy. They must be at least nine months old and weigh on average about 160 kg, having been fed with a diet that limits fat content to less than 10%. The hind legs used to produce the most expensive varieties of Ibérico ham, such as Bellota, come from animals fed exclusively on acorns.

[1] “Salt the hams. Put them whole into salt.”

Similar products exist in the cuisines of other nations as well. In Montenegro, a local variety of prosciutto called pršut is prepared; in southern France there is Bayonne ham; in Germany there are Black Forest, Mainz, Westphalian, and Ammerland hams. Spanish paleta is prepared in the same way as jamón, but from the swine's front leg. However, the method of preparation essentially remains the same. And the processes that take place during curing do not change simply because the hoof is left attached to the San Daniele ham leg, giving it its characteristic guitar-like shape.

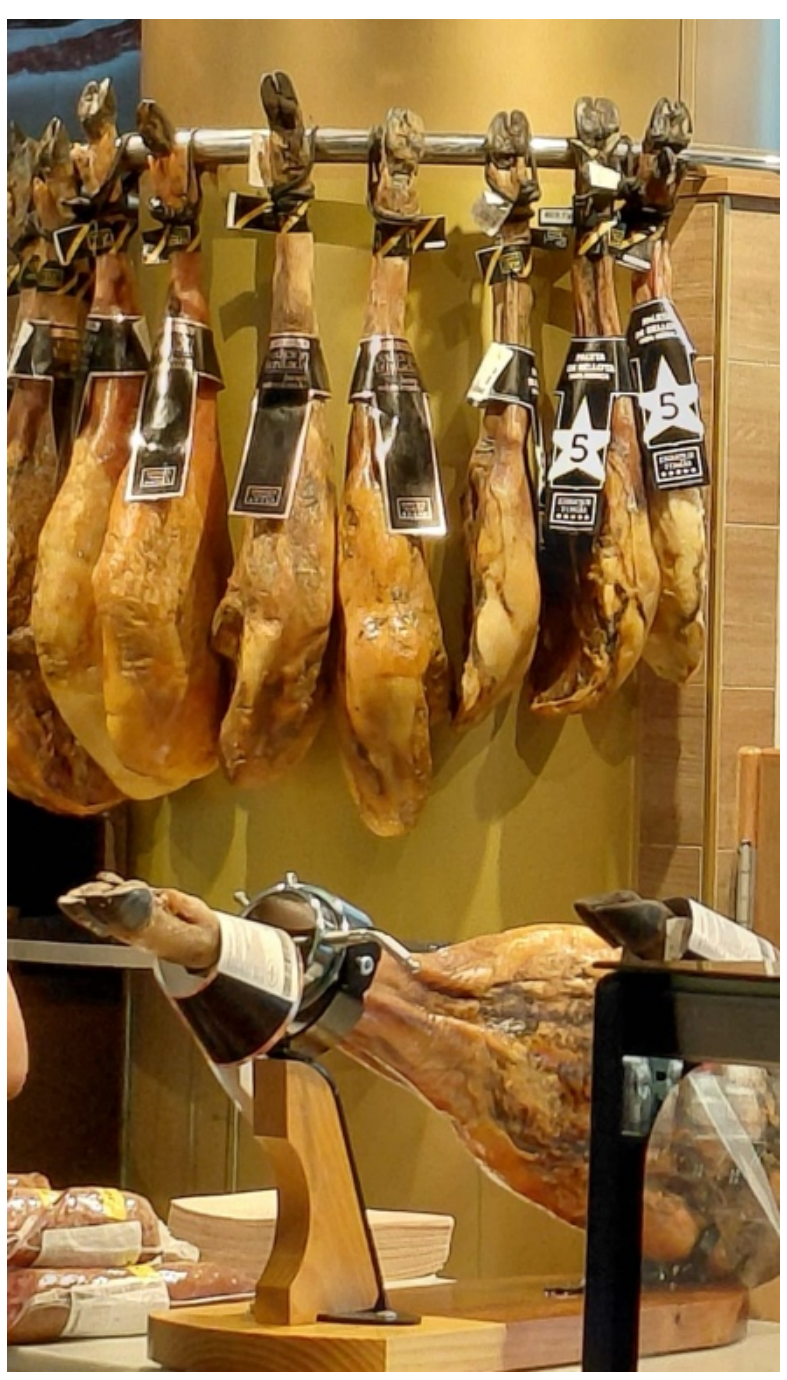

*Figure 2 Jamonería in San Sebastian (Spain). Couretesy of Giorgio Benedek*

### *Salting — one day in salt for each kilogram of weight*

Everything begins with dry salting: high-quality ham is generously rubbed with dry salt. This action triggers two different physical processes. The faster one is called osmosis, and the slower one is diffusion.

Meat consists of cells surrounded by membranes and outer layers. These are not very permeable to the sodium and chloride ions that form table salt, but they are permeable to the slightly smaller molecules of water[2]. Roughly speaking, they are semipermeable: water molecules can pass through such membranes, while salt ions are held back. As a result, water gradually moves outward through the membrane, moistening the salt. The process of osmosis continues until the concentration of water in the ham and in the salt layer becomes equal.

However, no one waits that long. The layer of damp salt is periodically removed and replaced with fresh salt. The moment for changing the salt comes when it has become wet and compacted, the crystals have dissolved into a sticky brine, and it no longer draws water out of the ham. In the classical method (Prosciutto di Parma / San Daniele), about 60–70% of the total salt is applied first, and after 7–10 days the soaked salt is removed. During the second salting, the remaining salt is applied for 10–18 days, after which it is also removed. The second portion of salt

---

[2] The size of a water molecule is 0.28 nm, while the sizes of hydrated ions—sodium ($Na^+$) and chloride ($Cl^-$)—are slightly larger, measuring 0.36 nm and 0.33 nm, respectively.

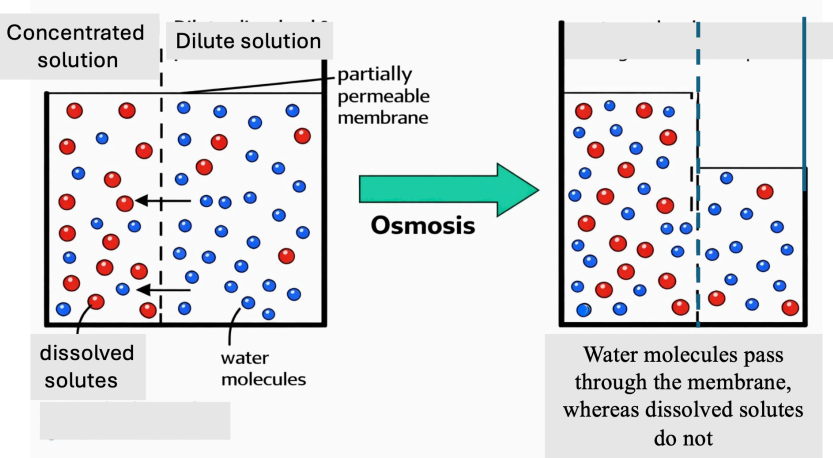

**Osmosis**

Osmosis is the process of spontaneous movement of water through a semipermeable membrane. Such a membrane can be an animal or plant tissue, a colloidal film, or similar materials. All of these membranes allow water to pass through but retain the substances dissolved in it.

Osmosis occurs because the membrane permits water molecules to pass but restricts the movement of dissolved particles, such as $Na^+$ and $Cl^-$. Water flows from the side of the membrane with a lower concentration $(C_l)$ of dissolved substances to the side with a higher concentration $(C_h)$ —in a sense, the solution “pulls” the solvent toward itself. This process continues until a certain equilibrium is established between the levels of water and the solution.

The origin of this phenomenon is straightforward to grasp. Because only water molecules can pass through the semipermeable membrane, equilibrium between the two vessels does not require equal total pressures on both sides. Instead, it is reached when the pressure in the vessel containing pure water matches the portion of the solution’s pressure contributed by its water molecules. As a result, the total pressure in the solution becomes higher than that in the pure water vessel by an amount attributable to the dissolved particles — this excess is known as the osmotic pressure of the solution.

Dutch physicist Jacobus Henricus van ’t Hoff described osmotic pressure—defined as the pressure required to prevent the dilution of a solute—as being directly related to the difference in concentrations on the two sides of a semipermeable membrane:

$$\Delta P_{osm} = (C_h - C_l) \cdot RT,$$

where $R = 8.31\, J/(mol \cdot K)$ is the universal gas constant, $T$ is the absolute temperature of the system.

The volumetric flux of water per unit membrane area, directed toward the region of lower pressure, can be described by the empirical relation (Weiss, 1996):

$$J = -L_p \Delta P_{osm}.$$

where $L_p$ is the hydraulic permeability of membrane. This parameter depends on the membrane’s specific composition and structure, which determine how easily water can pass through it.

One can notice its equivalence to the law of filtration

$$J = -\frac{\kappa}{\eta L} \Delta P.$$

First formulated in 1856 by the French engineer Henry Darcy during his work on water supply systems for the city of Dijon, this law shows us explicitly the structure of the coefficient relating the water flow with the properties of the filter. Here $\kappa$ is the porosity of the sand filling the filter, $L$ is the length of the filter, $\eta$ is the water viscosity.

becomes moist more slowly, and brine does not form even after the ham is turned over. In cold climates, particularly large hams may occasionally be salted a third time, although this is rare.

In the times of Ancient Rome, about 300–400 grams of salt per kilogram of meat were used, which made the ham extremely salty. Today, throughout the entire process, only 30–40 grams of salt per kilogram of meat are used for salting, preferably sea salt.

Only part of this amount ends up in the finished product; the rest is removed when the salt is brushed off the surface or drains away with the brine. The final salt content in Prosciutto di Parma and San Daniele is about 2–3%, while in jamón it is slightly higher—around 3–4%.

The duration of salting can be calculated using a simple rule: one day (or more) for each kilogram of meat. In some cases, this means up to a month. During this time, the air temperature should be 1–4 °C, with 75–80% humidity. Before the invention of refrigeration, producers tried to maintain a temperature of 2–6 °C, or in extreme cases up to 8 °C, so dry salting was carried out in cellars, usually from November to February. If salting began too early, when the weather was still warm, the meat could spoil. If it began too late, the salt might not have enough time to penetrate evenly, especially into the thickest parts of the ham. Modern producers use climate-controlled chambers and can therefore salt year-round. However, many adhere to tradition and salt only in winter.

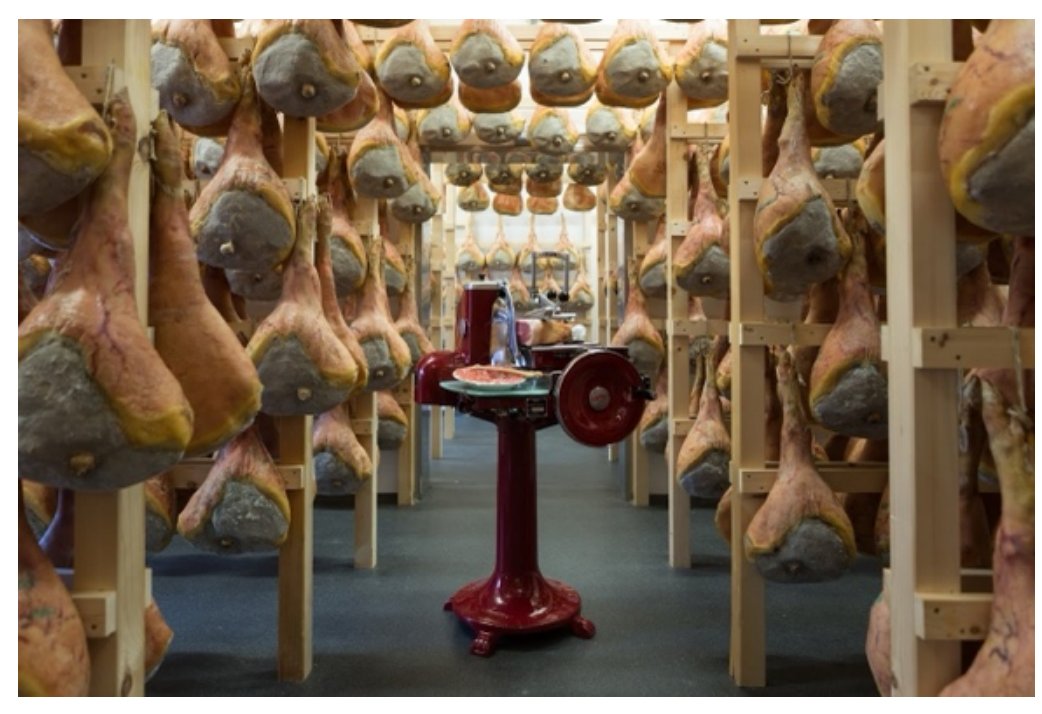

*Figure 3. The maturing of future prosciutto San Daniele in prosciuttoficio (ham factory) "La Glacere". The salted thighs are left to rest for up to four months from the start of processing in a special room with humidity varying between 70% and 80% and at a temperature between +4°C and +6°C. During this phase, the absorbed salt gradually penetrates and is distributed evenly. Courtesy of "La Glacere".*

What happens inside the ham during the salting process? First of all, the water content decreases, and the meat begins to dry[3]. The environment becomes unsuitable for the growth of pathogenic microorganisms such as *Clostridium botulinum*, *Listeria*, and *Salmonella*, although some of them or their spores may partially survive. During osmosis, water is also drawn out of the cells of harmful organisms, which leads to their death. Harmless bacteria, for example lactic acid bacteria, may survive and even participate in the subsequent fermentation of the meat, creating an acidic environment and displacing pathogens.

During this time, salt from the coating gradually penetrates into the ham through the process of diffusion. This process, which is slower than osmosis, involves the mutual movement of molecules of one substance among the molecules of another, leading to an equalization of their concentrations. In this case, salt (and spices, if they are used) slowly penetrates into the meat, increasing its saltiness.

At the early stage of salting, gravity also plays an important role. Under its influence, the brine flows downward; the lower part of the ham loses water faster and becomes more heavily salted[4]. For this reason, during the second salting the prosciutto is turned over, and a new layer of salt is applied unevenly, which helps reduce inhomogeneity. In jamón, this unevenness in saltiness may remain even in the final product. The ham is later hung up, which not only eliminates the effect of pools of brine but also stabilizes evaporation.

As a result of salting, the meat becomes partially dehydrated and salted, making it suitable for long-term preservation. But the process of making prosciutto does not end here. After salting, the meat must still "rest."

### *Riposo (Rest)*

The next stage is the air-drying (curing) of the resulting semi-finished product. This stage lasts from 10 months to 2 years and ends with the production of the fully matured product. During this long aging process, the ham loses most of its water through evaporation from the surface: prosciutto loses about 30–35% of its weight, while jamón loses 35–40%. Over this time, natural fermentation processes take place, and the ham develops its characteristic flavor. The standard for Prosciutto di Parma is a minimum curing period of 12 months, though it is often aged 18–24 months. During curing, enzymes break down proteins and fats, new flavor notes appear, and the color of the ham changes from pink to amber-red.

---

[3] Note that 30–40 g of salt per kilogram of meat cannot "draw out" all the excess water from the ham. Osmosis provides the initial rapid dehydration. Its purpose is to reduce the water content enough to prevent spoilage.

[4] Moist salt on the surface gradually slides down the ham under the influence of gravity. As a result, a thicker layer forms at the bottom, and the process of drawing out moisture occurs more intensively there.

Why does the evaporation process proceed so slowly? There is nothing technically difficult about speeding it up, but doing so would be a serious mistake. If evaporation happens too quickly, the surface becomes dehydrated: water evaporates faster than it can move from the center of the ham to the outside. As a result, the proteins on the surface denature and cross-link, which reduces the diffusion coefficient by an order of magnitude.

A common defect in curing is the formation of an undesirable crust on the surface of the ham, known as case hardening[5]. To prevent it, producers take all possible measures to control the rate of water evaporation from the surface. This rate is determined by three factors: temperature, humidity, and air velocity. Temperature is more or less fixed, but the other two factors—especially the speed of airflow—can be controlled. A light breeze should provide gentle convection rather than intensive drying. In modern prosciutto curing chambers the air speed is 0.05–0.2 m/s, which is 10–50 times lower than in industrial meat dryers, where the air speed is typically 1–3 m/s.

The proper air velocity can be determined not only experimentally but also through calculations. From the properties of meat, the flow of water toward the surface

**Diffusion**

If the concentration of a substance varies at different points in a system, then, thanks to the thermal motion of molecules, over time the dissolved substance moves from areas of higher concentration to areas of lower concentration until the concentration becomes uniform throughout the system. This process is called diffusion.

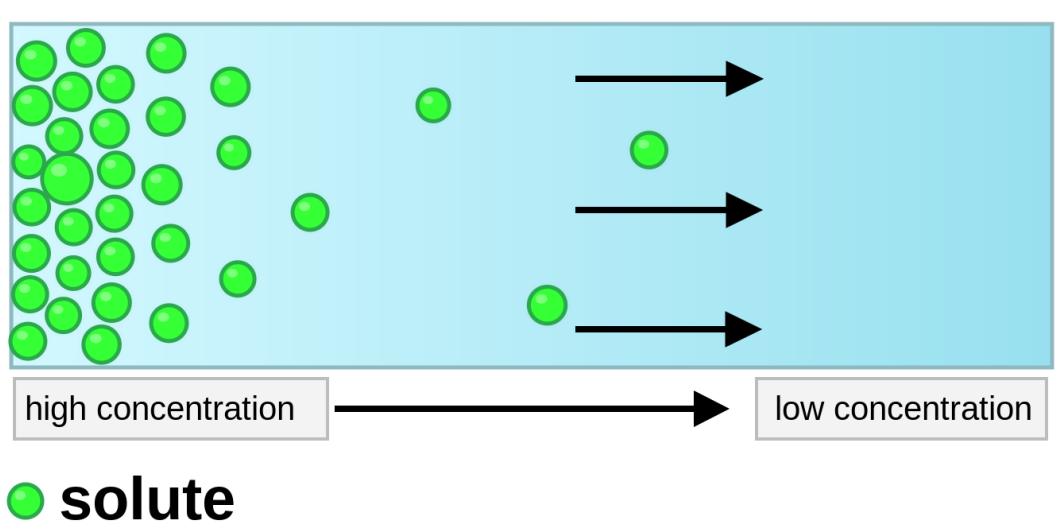

Let us assume that the volume concentration of the solution varies only in one direction, which we shall designate as the x-axis, and with time: $C \equiv C(x,t)$. Let us denote by the diffusion flux $J$ the amount of solute passing per unit time through a surface of unit area of the small box of cross-section $S$ and length d$x$: $J = {dN}/{Sdt}$, where the number of particles in the volume $dV = Sdx$, is $dN = C(x,t)Sdx$. Since the substance moves from regions of higher concentration to regions of lower concentration, the sign of the flux will be opposite to that of the derivative $dC/dx$. When the concentration of the solution is constant ($dC/dx = 0$), then there is no diffusion flux. Assuming $dC/dx$ to be not so large one can link the diffusion flux to the concentration gradient (first Fick's law):

$$J = -D\frac{dC(x,t)}{dx}.$$

where $D$ is a constant called the diffusion coefficient. Let us find the difference between the entering and exiting flows for the above-mentioned box of cross-section $S$ and length d$x$:

$$dJ = [J(x) - J(x+dx)] \approx -\frac{dJ(x,t)}{dx}dx = D\frac{\partial^2 C(x,t)}{\partial x^2}dx. \quad (*)$$

In light of the above definition of the flow, $dJ = \left({\partial C}/{\partial t}\right)dx$, and relation (*) can be rewritten in the form of the so-called second Fick's law:

$$\frac{\partial C(x,t)}{\partial t} = D\frac{\partial^2 C(x,t)}{\partial x^2}.$$

[5] Case hardening is a defect that occurs during the drying of products (meat, fruit, cheese), when the surface dries too quickly and forms a dense "crust" that prevents moisture from escaping from the interior.

lies in the range of $10^{-7}$ to $10^{-6}$ kg/($m^2$·s). Under typical temperature and humidity conditions, the evaporation rate is roughly a thousand times lower than the so-called mass transfer coefficient, which characterizes the air flow. This coefficient strongly depends on air velocity and is approximately proportional to the square root of the air speed. At 0.05 m/s, it is about 0.002 in SI units, which eliminates the risk of crust formation. At 0.1 m/s, the situation is borderline; if the speed increases threefold, a crust begins to form. It will almost certainly form at airflow speeds above 0.5 m/s. In real curing chambers, airflow is kept even lower, taking into account both its uneven distribution and the fact that evaporation from the surface is further slowed by fat and by the sugna coating, which we will discuss shortly. For prosciutto, the optimal air speed is considered 0.05–0.15 m/s, while for the thicker jamón it is slightly lower—about 0.03–0.1 m/s.

Naturally, an even more important parameter is air humidity. At 85% humidity, the ham loses water slowly and shrinks evenly, but already at 75% humidity a surface crust may form. In addition, the meat dries under these conditions but does not mature as quickly (we will return to this point later). This determines the choice of optimal humidity. For the thicker ham used in jamón, with its large bone, a humidity of about 85% during the first months is desirable; otherwise the airflow disrupts the balance of diffusion and a crust forms. Iberian ham cracks less often because of its higher fat content.

After 3–4 months of drying, the ham is coated with a mixture of pork fat, wheat or rice flour, and sometimes black pepper and salt, called sugna. This coating slows down drying and helps prevent cracks. Coatings of different colors can be seen in Figure 1.

There is also an additional nuance that is rarely mentioned. Mold appears on the surface of the ham, and it is removed only at the very end of the process. At a humidity of 80–85%, this is typically stable Penicillium, which is beneficial. It protects the surface, prevents oxidation, competes with pathogenic microflora, stabilizes moisture, and contributes mild mushroom and nutty notes to the flavor.

Unlike fermented sausages such as salami or French cheeses like Brie, Roquefort, or Camembert, the correct mold on ham is not so much an active participant in the maturation process as it is a protective surface layer. It is rarely introduced deliberately as a starter culture; instead, it usually grows naturally. If the humidity becomes too high, there is a risk of undesirable strains appearing; these are washed off with vinegar or wine. If the humidity is lower, mold tends to grow in patches on the drier surface. In all cases, the mold does not penetrate deep into the meat. It influences the flavor slightly, but not significantly. In the finished product, no mold remains.

Thus, for a number of reasons, the airflow must be slow and the humidity relatively high. This is easy to achieve in modern curing chambers. In the past, in Italy, prosciutto was traditionally dried in ventilated stone cellars or attics with adjustable shutters, using the natural mountain breeze. The windows were small and usually faced the sea. In the area around Parma, they faced toward the Apuan Alps and the Ligurian Sea beyond them; in San Daniele, on the eastern slopes of Friuli, they faced toward the Adriatic Sea.

### *Sliced for Connoisseurs*

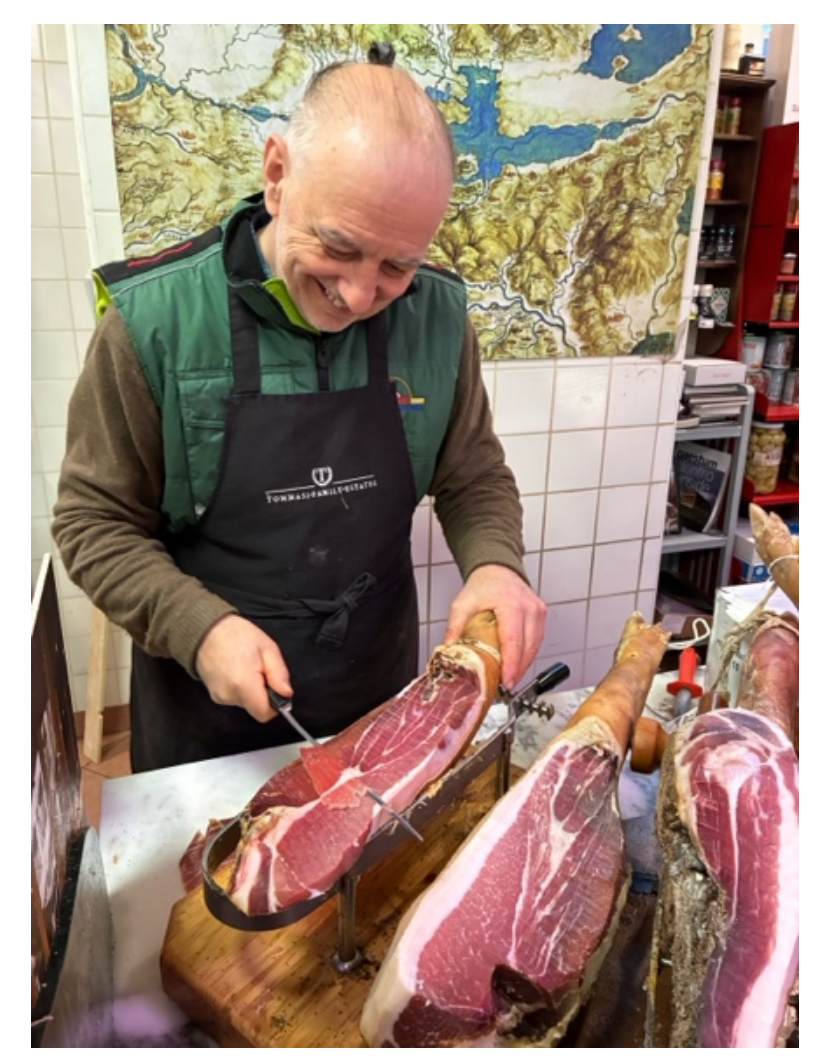

Figure 4 Signor Claudio Lunghini slices prosciutto Toscano.

When visiting a prosciutto shop, a connoisseur assesses the appearance of the various cuts of prosciutto on display, paying attention to their color, fat content and its distribution. He also examines the cross-section of the piece on offer and discusses with the seller the characteristics of each piece — whether its flavor leans towards sweet or salty, its curing time and region of origin.

Finaly, the last operation is required: it must be cut into extremely thin slices, and in a particular way. In modern supermarkets this is usually done with a special cutting machine (slicer). However, with this method the prosciutto can heat up because of friction between the blade and the meat, which in a matter of seconds may lead to protein denaturation—something that the producers of prosciutto carefully avoided during the many months of its preparation. For this reason, it is recommended to slice it at low speed[6].

Even so, it is much better when prosciutto is cut by hand in your presence, using a long, thin, and very sharp knife (see Fig. 4). Not every vendor masters this skill, and the slices are often not as thin and perfectly even as those produced by machine slicing. However, the unsurpassed flavor of prosciutto, with its delicate aroma and texture, is preserved much better this way.

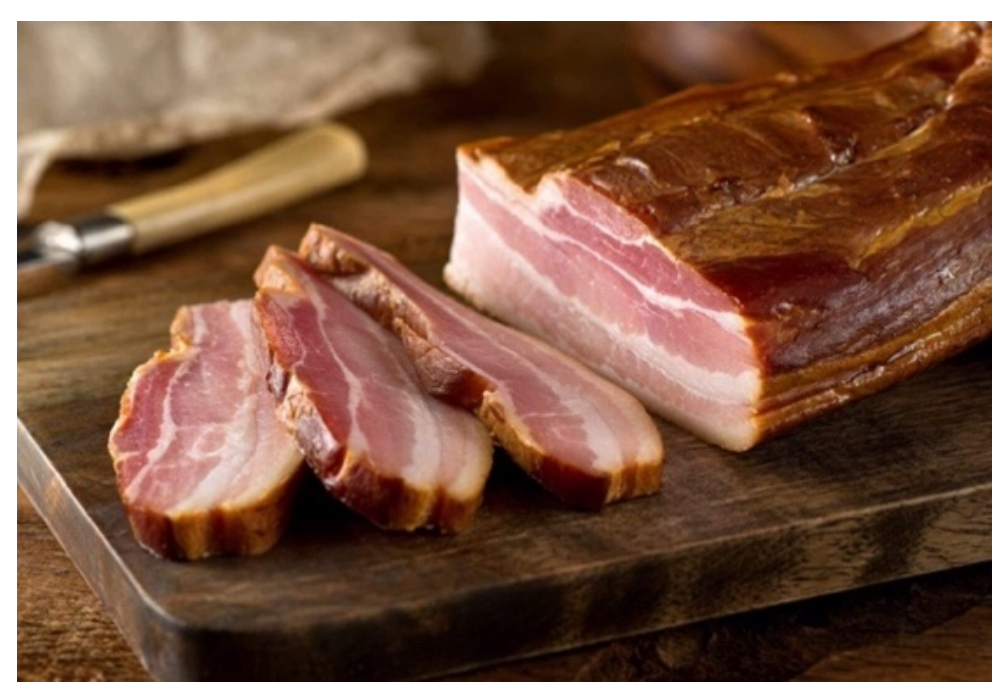

Figure 5 Italian pancetta.

### *Bacon and pancetta*

A procedure similar to prosciutto salting, but much simpler, can be seen in the preparation of bacon—a far cheaper pork product (in the Italian version, this is pancetta). It is usually made from pork belly, and less often from the back.

In dry salting, meat is rubbed with a mixture of salt and sugar to preserve its color and prevent bacterial growth. Sometimes, sodium nitrite is added; although it is not particularly healthy, it increases shelf life. In wet salting, the meat is soaked in a brine. The meat is left in the refrigerator for several days to a week to allow it to absorb the salt.

[6] Meat, basically, consists of complex organic macromolecules called proteins (the type of protein varies from the type of meat). In raw meat, protein molecules are in a state of entangled long chains. In the course of heat treatment, the temperature rises, these chains straighten, and when the temperature reaches the specific for each type of meat value, they are compactified into some kind of "carpet". This process is called protein denaturation. It occurs at relatively low temperatures: for meat it is 55~80°C, for fish the temperature is even lower.  The entire process of transforming meat into prosciutto is designed to avoid even the slightest hint of denaturation, which would greatly degrade the flavor of the product.

The resulting product is sometimes smoked, which increases its shelf life and changes its flavor. Typically, the salting times for meat intended for hot or cold smoking are: Pork: 5–7 days; Beef: 7–10 days; Poultry: 2–3 days.

### ***Some Estimates***

Suppose we are making prosciutto from a pork ham weighing 18 kg. About 4–5 kg will be bone, and the remaining 13–14 kg will be meat. The ham is 50–60 cm long, 25–30 cm wide at its widest point, and 15–20 cm thick at its thickest point. Let's illustrate the qualitative discussion with numerical estimates.

The first question concerns diffusion: how deeply do salt molecules penetrate during the salting time $t$? The answer is given by Fick's Second Law for one-dimensional diffusion. Its solution in this case takes the form:

$$C(x,t) = \frac{C_S(0,0)}{\sqrt{4\pi Dt}} exp\left(-\frac{x^2}{4Dt}\right),$$

where $C(x,t)$ is the volume concentration of the salt at the distance $x$ from the plane surface, $C_S(0,0)$ is the surface concentration of the salt at the plane $x = 0$, when at the moment $t = 0$ the salt was placed there, $D$ is its diffusion coefficient of the salt in meat[7].

As can be seen from this formula, the characteristic salt penetration depth, at which its concentration decreases by a factor of $e/\sqrt{\pi} = 1.53$, is

$$x_c = 2\sqrt{Dt}\,.$$

The diffusion coefficient of salt in meat D depends on the type of meat, the curing method, and the temperature. We use the value $6\times10^{-10}$ m$^2$/s [8]. Let's convert it to more convenient units of time. Salting time is measured not in seconds, but in days. Considering that 1 day = 86,400 seconds, we get

$$x_c = 14.4\,mm\sqrt{\frac{t}{1\,day}}\,.$$

For our ham, this is 20–60 days, which gives an estimate of the salt penetration depth $x_c$ of 6.4 to 11 cm. This is less than the full size of the ham. Yet, prosciutto purchased in a store is edible throughout the entire cross-section.

---

[7] L.D. Landau, E.M. Lifshitz, *Hydrodynamics*, section 59.

[8] Christian Vestergaard et al, "Sodium diffusion in cured pork determined by $^{22}$Na radiology" Meat Science, **76**(2), 258 (2007); Christian L. Hansen et al, "Diffusion of NaCl in meat studied by $^{1}$H and $^{23}$Na magnetic resonance imaging", Meat Science, **80**(3), 851 (2008).

The reason is that although, after transitioning from the salting phase to drying (riposo), the influx of salt from the surface stops, the diffusion process continues. Salt inside the ham keeps redistributing itself, seeking to equalize its concentration in different parts.

Usually, instead of using Fick's law, a simplified empirical estimate is applied, based on an average salt propagation rate of ~1–2 mm/day depending on conditions. According to this estimate, salt will penetrate 3–12 cm during the 20–60 days salting period, and during the drying phase (300–700 days), it will evenly season the entire ham.

### *Beyond the Latin World: Pastirma*

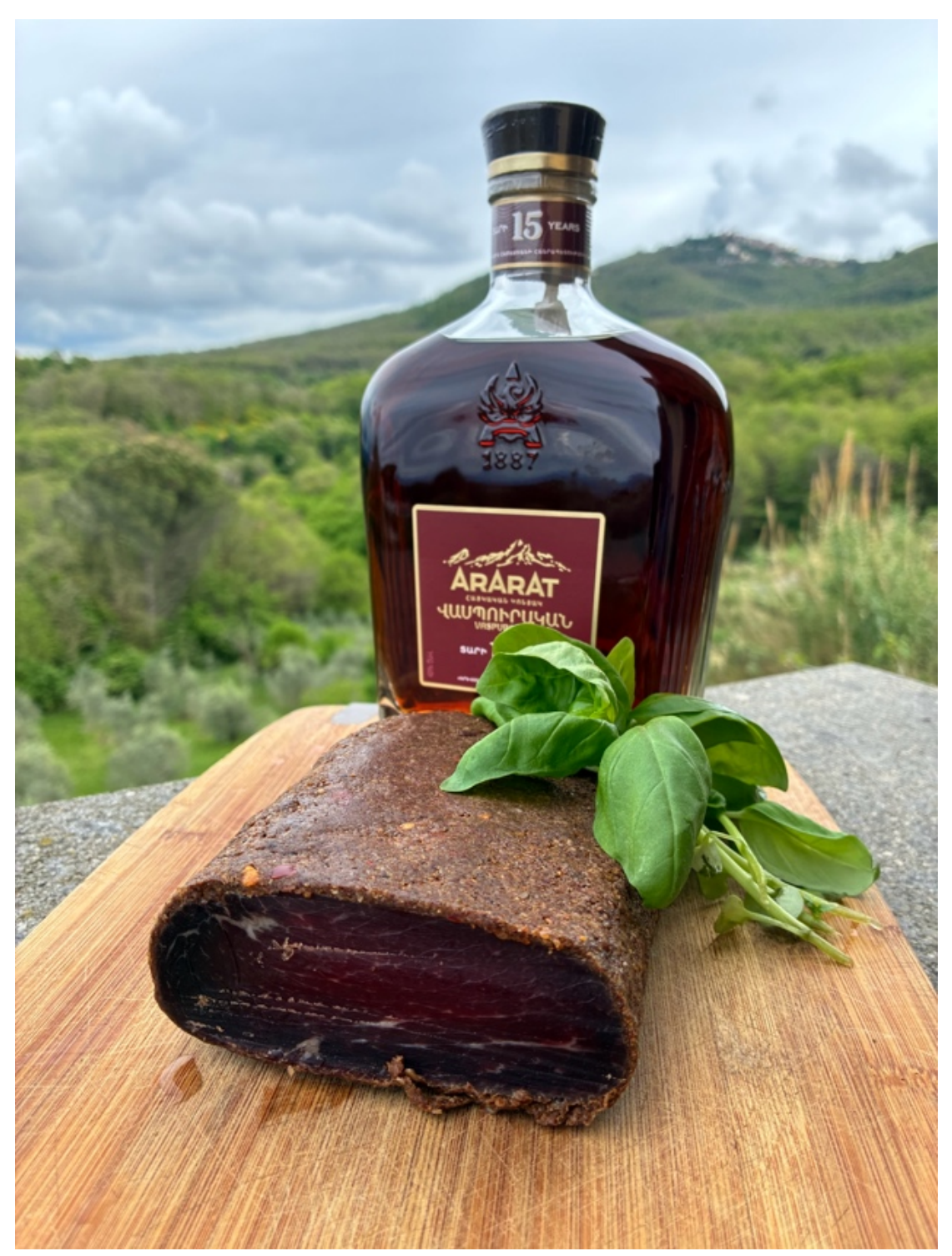

*Figure 6 Armenian basturma. Courtesy of Robert Vardanyan.*

Who says that water cannot be removed from meat in a fundamentally different way than in the preparation of prosciutto? The record for protein content belongs to *pastirma*[9]. It is a heavily salted meat coated with a spice paste, typically made from beef, and is part of the culinary traditions of the Caucasus and the Near East. Pastirma is thinly sliced and served as an appetizer. The distinctive feature of pastirma preparation is, following butchering, dry salting, and rinsing, the pressing of the meat. The meat is placed under a weight for 1–3 days, which replaces a month-long drying process and gives it a flat shape. The pressed meat is then dried, coated with spices, and further dried. Rapid dehydration through the expulsion of intercellular water is complemented by a powerful spicy barrier.

Whole long cuts of beef are used (less often pork or other meats) with low fat content. Parallel muscle fibers help ensure even salt diffusion. The meat is generously coated with salt and held at a slight angle for 5–10 days at around 4–8 °C: osmosis draws out the water, and the resulting brine flows downward. The salt is then rinsed off, and sometimes the meat is briefly soaked. This reduces surface oversalting, though enough salt has already penetrated the interior. After rinsing, reverse diffusion occurs: part of the salt returns to a thin layer (about a millimeter) near the surface. Finished pastirma usually contains 4–6% NaCl, making it saltier than prosciutto or jamón.

[9] Pastirma (also spelled pasturma, bastirma, or basturma) is air-dried, cured beef.

After pressing, the meat is pre-dried in air for 2–5 days to remove residual moisture and stabilize its shape. It is then coated with çaman—a spice mixture with fenugreek. Classic ingredients include garlic, paprika, and cumin seeds, with occasional additions of hot pepper, coriander, and nutmeg. Çaman not only adds flavor but also slows evaporation, prevents case hardening, inhibits microbial and mold growth (unlike prosciutto, mold is undesirable here), and acts as an antioxidant. The final drying lasts 7–20 days at moderate temperature and 60–75% humidity. During this time, partial salt-induced protein denaturation occurs; there is no prolonged "aging." Flavor is primarily formed by salt and spices, with minimal fat oxidation. Thanks to pressing, the entire process takes weeks rather than years, and the finished pastirma has a dense texture.

Beef is also air-dried in Italian cuisine, producing, for example, bresaola, but this process is closer to prosciutto preparation and does not involve pressing. Enzymes play a more important role in bresaola, giving it a more delicate flavor than pastirma. Bresaola takes longer to make but avoids the coarse mechanical force of pressing — "soft force" versus "Sturm und Drang".

As you can see, gourmets have plenty to choose from… At the same time, it is worth remembering that in the preparation of all these products, two physical phenomena are involved: osmosis and diffusion, in addition to chemical and biological transformations. Understanding these mechanisms allows one to perfect the ancient art of creating melting-in-the-mouth cured ham.

The Roman sonneteer Giuseppe Gioachino Belli once wrote: *"Er pane, er vino, e 'na fetta de prosciutto fanno scordà li guai d'un giorno brutto"* [10]. A slice of prosciutto, he suggests, is enough to make one forget the troubles of a bad day. For physicists, however, such a slice offers more than simple consolation: while savoring it, they may also glimpse the subtle physical laws that underlie this delicacy of Mediterranean cuisine.

The authors express their deep gratitude to Sergey Artemenko, Giorgio Benedek, Michail Chernikov, Giuseppe Eramo, Claudio Lunghini, Claudia Poggioni, Sergey Salikhov, Yuriy Yerin, and Robert Vardanyan for their valuable comments.

[10] *"*Bread, wine and a good ham slice may
help you forget the troubles of the day*"*